\documentclass{aa} 

\usepackage{natbib}
\bibpunct{(}{)}{;}{a}{}{,} 

\usepackage{graphicx}

\usepackage{amssymb,amsmath}
\usepackage{delarray}
\usepackage{mathrsfs}
\usepackage{xcolor}

\usepackage[varg]{txfonts}
\usepackage{float}

\usepackage{soul} 

\begin{document} 

\title{Observational indications of magneto-optical effects in the\\ 
scattering polarization wings of the Ca~{\sc i} 4227\,\AA\ line}

\author{Emilia Capozzi, \inst{1}
          Ernest Alsina Ballester, \inst{1}
          Luca Belluzzi, \inst{1,}\inst{2}
          Michele Bianda, \inst{1}
          Sajal Kumar Dhara, \inst{1}\\
		  and
          Renzo Ramelli \inst{1}
		  }

\institute{Istituto Ricerche Solari Locarno, 6605 Locarno Monti, Switzerland\\  
\email{emilia.capozzi@irsol.ch}
\and Leibniz-Institut f\"ur Sonnenphysik (KIS), 79104 Freiburg, Germany
}

\titlerunning{Observational indications of magneto-optical effects 
in Ca~{\sc i} 4227\,{\AA}}
\authorrunning{Capozzi et al.}

\abstract
{Several strong resonance lines, such as H~{\sc i} 
Ly-$\alpha$, Mg~{\sc ii} k, Ca~{\sc ii} K, Ca~{\sc i} 4227\,\AA, 
which are characterized by deep and broad absorption 
profiles in the solar intensity spectrum, 
show conspicuous linear scattering polarization signals
when observed in quiet regions close to the solar limb.
Such signals show a characteristic triplet-peak 
structure, with a sharp peak in the line core and extended wing lobes.
The line core peak is sensitive to the presence of magnetic fields through the 
Hanle effect, which however is known not to operate in the line wings.
Recent theoretical studies indicate that, contrary to what was previously believed, the wing linear
polarization signals are also sensitive to the 
magnetic field through magneto-optical effects.}
{We search for observational indications of this recently discovered physical mechanism in the 
scattering polarization wings of the Ca~{\sc i} 4227\,\AA\ line.}
{We performed a series of spectropolarimetric observations of this line using 
the Zurich IMaging POLarimeter (ZIMPOL) camera at the Gregory-Coud\'e telescope of IRSOL (Switzerland) and 
at the GREGOR telescope in Tenerife (Spain).}
{Spatial variations of the total linear polarization degree and of the linear polarization angle 
are clearly appreciable in the wings of the observed line. 
We provide a detailed discussion of our observational results, showing that 
the detected variations always take place in regions where longitudinal 
magnetic fields are present, thus supporting the theoretical prediction that they are produced by MO effects.}
{}


\maketitle

\section{Introduction}\label{sec:sec1}
Over a decade ago, unexpected observations of the scattering 
polarization signal of the Ca~{\sc i} 4227\,\AA\ line were reported 
\citep[see][]{Bianda03}.
This signal shows a triplet-peak structure, with a sharp peak 
in the line core, and broad lobes in the wings. The wing lobes, which are 
typical of strong resonance lines, are produced by coherent scattering 
processes.
The observations of \citet{Bianda03} were unexpected because, in addition to spatial variations in the Stokes $Q/I$ wing 
signal (the reference direction for positive $Q$ being the parallel to the closest limb), $U/I$ wing
signals of substantial amplitude and showing similar spatial variations were found.
While spatial variations of the $Q/I$ wing lobes could in 
principle be explained in terms of variations of the anisotropy of the 
radiation field, or of particular properties of the atmospheric plasma (e.g., the 
density of neutral perturbers, which determine the fraction of coherent 
scattering processes), no straightforward physical explanation 
was readily available for the appearance and variation of $U/I$ wing signals.

The observation of $U/I$ signals in the line wings could not 
be explained in terms of the Hanle effect, as this is known to operate in the 
core of spectral lines and not in the wings \citep[for an 
exhaustive overview, see][]{Stenflo82,TB01,LandiE}.
This circumstance actually lead to the common belief that the large wing 
scattering polarization signals shown by strong resonance lines were 
insensitive to the presence of magnetic fields.
On the other hand, \citet{Bianda03} reported that the observed 
variations in the $Q/I$ and $U/I$ wings of the Ca~{\sc i} 4227\,{\AA} line 
were always accompanied by Zeeman signals in the nearby blends, thus suggesting that 
such unexpected wing behavior could indeed have a not yet understood magnetic origin.
\citet{Sampoorna09} investigated the possibility that the Hanle effect could extend in the wings,
when a high rate of collisions and strong magnetic fields are present. However they ruled out
this explanation when considering realistic collisional rates and field strengths.
We finally observe that at the spatial and temporal resolutions of the observations 
of \citet{Bianda03} it can be excluded that the detected $U/I$ signals could be produced by horizontal 
inhomogeneities in the solar plasma.

Theoretical investigations carried out during the last few years \citep[see][]{DelPinoAleman2016,Alsina16}
have shown that, in contrast to what was previously thought, the conspicuous 
scattering polarization signals observed in the wings of many strong resonance lines should be sensitive to the presence of 
magnetic fields.
This sensitivity arises when longitudinal magnetic fields are present due to 
the so-called magneto-optical (MO) effects.
These effects most notably induce a rotation of the plane of linear polarization
\citep[often referred to as Faraday rotation, e.g.,][]{Pershan67},
modifying the $Q/I$ wing signal and giving rise to an appreciable wing 
signal in $U/I$. 
MO effects have been theoretically predicted to have an appreciable impact on 
the scattering polarization wings of a number of strong resonance lines, 
including the Mg~{\sc ii} k \& h lines \citep{DelPinoAleman2016,Alsina16}, 
the Sr~{\sc ii} line at 4078\,\AA\ \citep{Alsina17}, the H~{\sc i} Ly-$\alpha$ 
line \citep{Alsina19}, and the Ca~{\sc i} line \citep{Alsina18}.
These theoretical investigations suggest MO effects as an ideal candidate for 
explaining observations like those of \citet{Bianda03}.
Moreover, they could offer a novel window into the magnetism 
of the solar atmosphere, providing complementary information to that accessible through the Hanle and Zeeman effects.

In this work we present new spectropolarimetric observations
of the Ca~{\sc i} 4227\,{\AA} line, carried out in regions having different 
levels of magnetic activity. 
We provide indications that variations of the $Q/I$ wing lobes, together 
with the appearance of significant $U/I$ wing signals, are observed in 
regions where relatively strong longitudinal fields are present, thus 
supporting the theoretical prediction that such variations are produced by MO 
effects.
In section \ref{sec:sec2} we analyze the scattering polarization signal of the Ca~{\sc i} 4227\,{\AA} line, 
and we introduce the main physical quantities on which we will base our analysis.
In sections \ref{sec:sec3}, \ref{sec:sec4} and \ref{sec:sec5} we present 
our spectropolarimetric observations, which were carried out with the Zurich Imaging Polarimeter 
(ZIMPOL), both at GREGOR and at IRSOL. 
For each of them, we briefly describe the instrumental set-up, we provide 
information on the observed solar region, and we describe the data reduction 
procedure.
Finally, we analyze the data, and we provide our conclusions.

\section{The scattering polarization signal of the Ca~{\sc i} 4227\,\AA\ 
line}\label{sec:sec2}

\begin{figure}[t!]
\centering
\includegraphics[width=.5\textwidth]{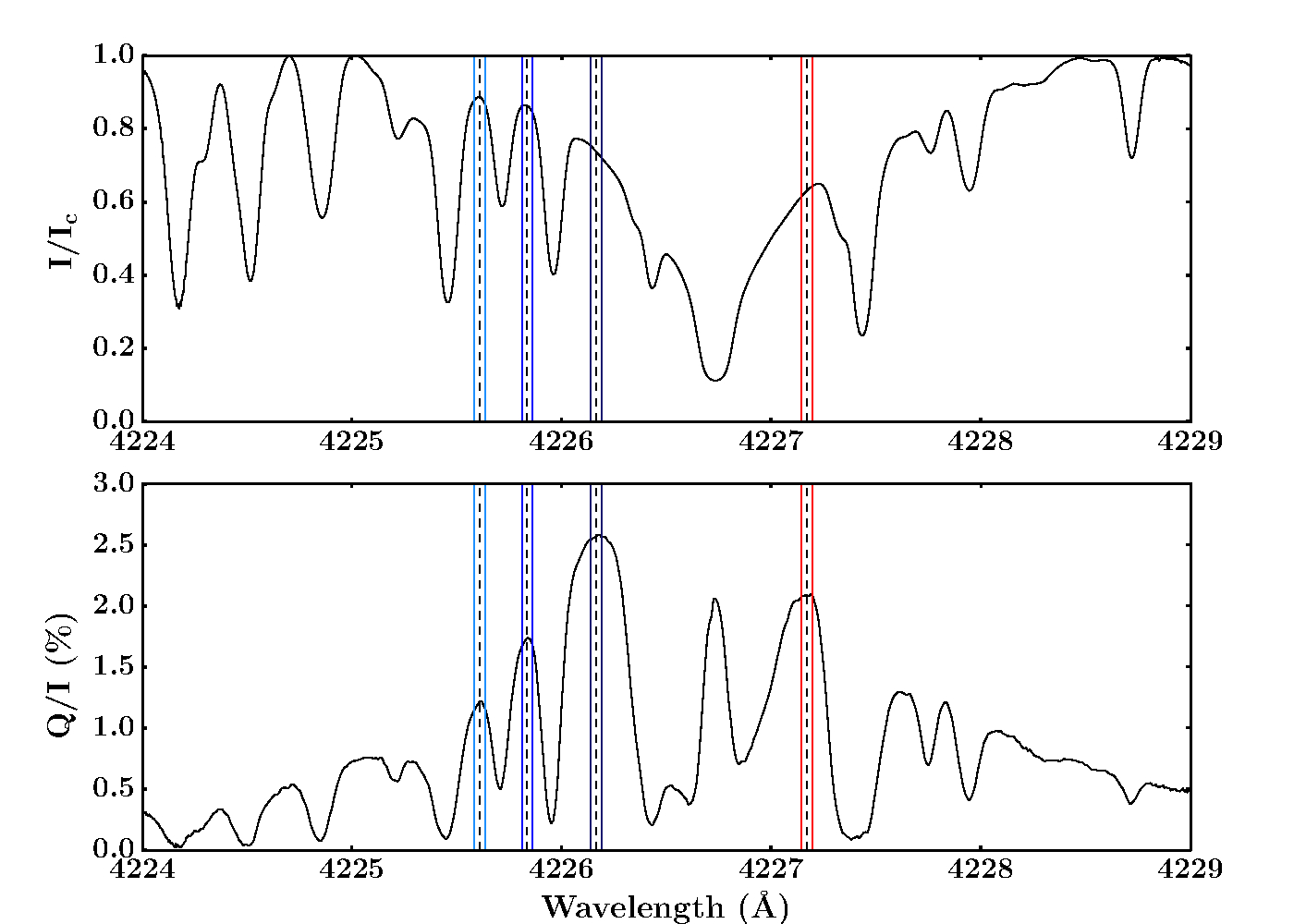}
	\caption{Intensity spectrum (upper panel) and Second Solar Spectrum 
	(lower panel) across a wavelength interval containing the Ca~{\sc i}
	4227\,\AA\ line \citep[data from][]{Gandorfer02}. 
	The blue and red lines indicate the wavelength intervals of interest for 
	this work (see text).}
\label{fig:atlasrangewings}
\end{figure}

\begin{figure}[t!]
\centering
\includegraphics[width=.45\textwidth]{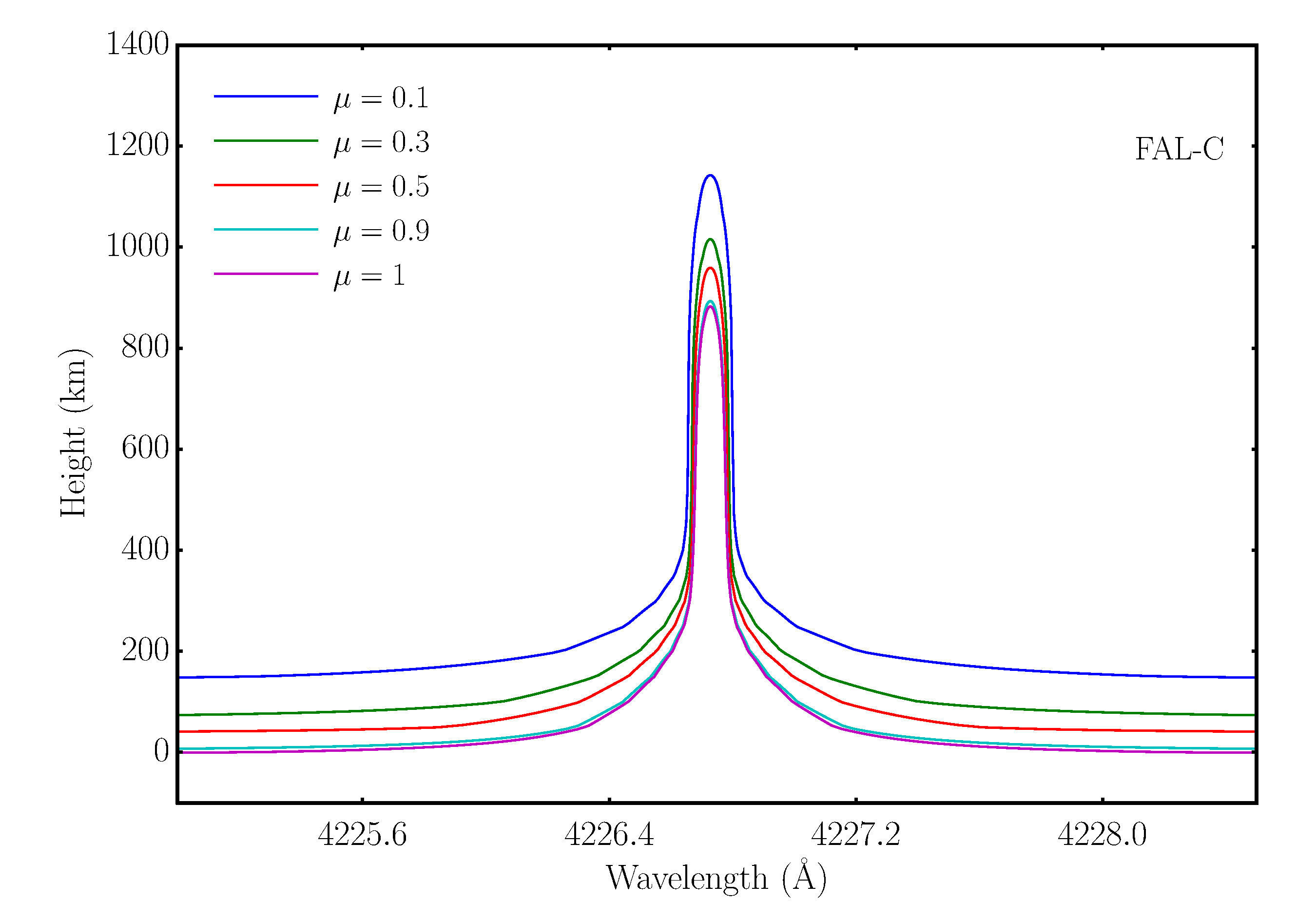}
\caption{Height at which the optical depth is unity as a function of wavelength 
	across the Ca~{\sc i} 4227\,\AA\ line, in the semi-empirical atmospheric 
	model C of \cite{Fontenla93}, for various lines of sight (see legend).
	These curves provide a rough estimate of the formation region of the line 
	as a function of wavelength.} 
\label{fig:formationheight}
\end{figure}

In the solar intensity spectrum, the Ca~{\sc i} 4227\,\AA~line is characterized 
by a broad absorption profile with extended wings (see the upper panel of 
Figure~\ref{fig:atlasrangewings}). 
It is produced by a resonance transition between the ground level 
of neutral calcium, which has total angular momentum 
$J_\ell=0$ [$^{1}$S$_{0}$], and an upper level with 
$J_{u}=1$ [$^{1}$P$^{\rm o}_{1}$].
The line-core radiation originates primarily from atmospheric heights around 
1000~km above the $\tau=1$ surface, making it chromospheric in nature 
\citep{Supriya}, while the radiation at wavelengths immediately outside the 
core of the line originates from photospheric heights, as illustrated in 
Figure~\ref{fig:formationheight}.

This line exhibits the largest scattering polarization signal in the visible 
range of the so-called Second Solar Spectrum 
\citep{Bruckner63,Stenflo80,Stenflo84,Gandorfer02}. 
The $Q/I$ profile is characterized by a sharp peak in the 
line core and prominent lobes in the wings (see the lower panel of 
Figure~\ref{fig:atlasrangewings}). 
This characteristic triplet-peak structure is typical of 
strong resonance lines. The $Q/I$ wing lobes arise as
a consequence of frequency coherent scattering processes
\citep{Holzreuter2005}. 
As such, it can be shown that their amplitude is insensitive 
to the depolarizing effect of elastic collisions with neutral perturbers. 
On the other hand, elastic collisions may modify the amplitude
of the scattering polarization wings through the branching ratios determining 
the fraction of coherent and non-coherent scattering processes
\citep[see][]{Alsina18}.
The drops in amplitude that can be observed within the wing lobes of the $Q/I$ 
signal result from the presence of several blended lines, especially of 
Fe~{\sc i}, which can be clearly observed also in the intensity spectrum.

In this work we search for observational evidences of the 
theoretically predicted sensitivity of the scattering polarization wing lobes 
of the Ca~{\sc i} 4227\,\AA\ line to MO effects by comparing 
the spatial variations of the linear polarization signals to the magnetic 
activity within the same regions.
To this aim, we analyze the linear polarization angle (i.e., the angle between 
the direction of linear polarization and the specified reference direction)
and the total linear polarization fraction at particular wing wavelengths. 
The polarization angle, which is defined within the interval 
$[0^\circ,180^\circ)$ \citep[e.g.,][]{LandiE}, is given by 
\begin{align} 
\label{eqalpha} 
\alpha  &= \frac{1}{2} \tan^{-1} \left(\frac{U}{Q}\right) + \alpha_{0}, \\
\text{with}\ \ \  
\alpha_{0} &= 
\begin{cases}
0 ^\circ,  \ \ \ \ \ \text{if} \ Q > 0 \ \text{and} \ U \geq 0 \\
180 ^\circ,\ \text{if} \ Q > 0 \ \text{and} \ U < 0 \quad . \\
90 ^\circ,\ \ \ \text{if} \ Q < 0
\end{cases}
\end{align}
According to this definition, when $U/I = 0$, $\alpha = 0^\circ$ 
if $Q/I > 0$ and $\alpha = 90^\circ$ if $Q/I < 0$, while when $Q/I = 0$, 
$\alpha=45^\circ$ if $U/I > 0$, and $\alpha=135^\circ$ if $U/I < 0$.
The linear polarization angle is of course undefined when $Q = U = 0$.
The total linear polarization fraction is given by:
\begin{equation}\label{eqpl}
P_{L}= \sqrt{(Q/I)^2+(U/I)^2}.
\end{equation}
We note that the total linear polarization fraction of scattering polarization signals
depends directly on the degree of anisotropy of the scattered radiation field.

The MO effects operating in the presence of a magnetic field with a 
longitudinal component produce a rotation of the plane of linear polarization, 
giving rise to a non-zero $\alpha$ angle. On the other hand, 
one may in principle expect such MO effects to preserve the polarization 
fraction $P_L$.
However, it should be noted that the emerging radiation is generally emitted 
over an extended region in the solar atmosphere.
Since the rotation of the plane of linear polarization depends on the distance 
traveled through the medium, radiation emitted at different points is rotated 
by different amounts.
As a consequence, the total linear polarization fraction of the radiation can 
decrease \citep[see the Appendix A in][]{Alsina18}.

In order to reduce the influence of blends as much as possible, we focus our 
attention on the wing wavelengths at which the $Q/I$ signal shows local maxima 
(peaks). 
We consider four different wavelengths, three 
in the blue wing (moving from the core to the wing, they will be referred to 
as bI, bII, and bIII) and one in the red wing (hereafter referred to as rI).
Due to the noise affecting the observed profiles, the position of the $Q/I$ 
peaks is determined by performing a Gaussian fit of the observed signal around 
the considered peak, and by finding the maximum of the ensuing curve.
To minimize the noise in the physical quantities that will be analyzed 
($\alpha$ and $P_L$), the observed signal is averaged over a small spectral 
interval centered at the fit maximum.
The width of this interval must be large enough to reach an acceptable
signal-to-noise ratio, but sufficiently narrow to not include 
spectral regions affected by the blends.
We have found a width of 50\,m{\AA} to be suitable. 
We have verified that the values of $\alpha$ and $P_L$ do not change 
appreciably if this value is slightly modified.
Taking the observation of \citet{Gandorfer02} as reference, the 
wing intervals, determined as described above, are shown in 
Figure~\ref{fig:atlasrangewings}. 
Such wing intervals can be specified through their distance from the line 
center, determined by finding the intensity absorption profile minimum, 
following a fitting procedure fully analogous to the one previously described.
In all the observations considered in this work, the wavelength intervals of 
interest are determined from the $Q/I$ profiles averaged over the entire slit, 
following the aforementioned workflow.

\section{The Zeeman $V/I$ signal in the nearby Fe~{\sc i} 4224.2\,{\AA} 
line}\label{sec:sec3}
For the purpose of determining the magnetic origin of the 
spatial variations of the wing scattering polarization of the Ca~{\sc i} 4227\,\AA~line, 
we analyze the circular polarization signal produced by the Zeeman effect in 
the nearby Fe~{\sc i} 4224.2\,\AA\ line. 
Although it must be acknowledged that in general Stokes $V$ 
is not strictly proportional to the longitudinal component of the magnetic field \citep[see][]{inversion}, it 
nevertheless provides a reliable indication of the level of magnetic activity of the observed region.
For all the measurements presented in the following sections, we have considered the maximum amplitude 
of the blue peak of the $V/I$ signal of the aforementioned Fe~{\sc i} line, because the red peak is closer 
to another blended line.

\begin{figure*}[t!]
\centering
\includegraphics[width=1.\textwidth]{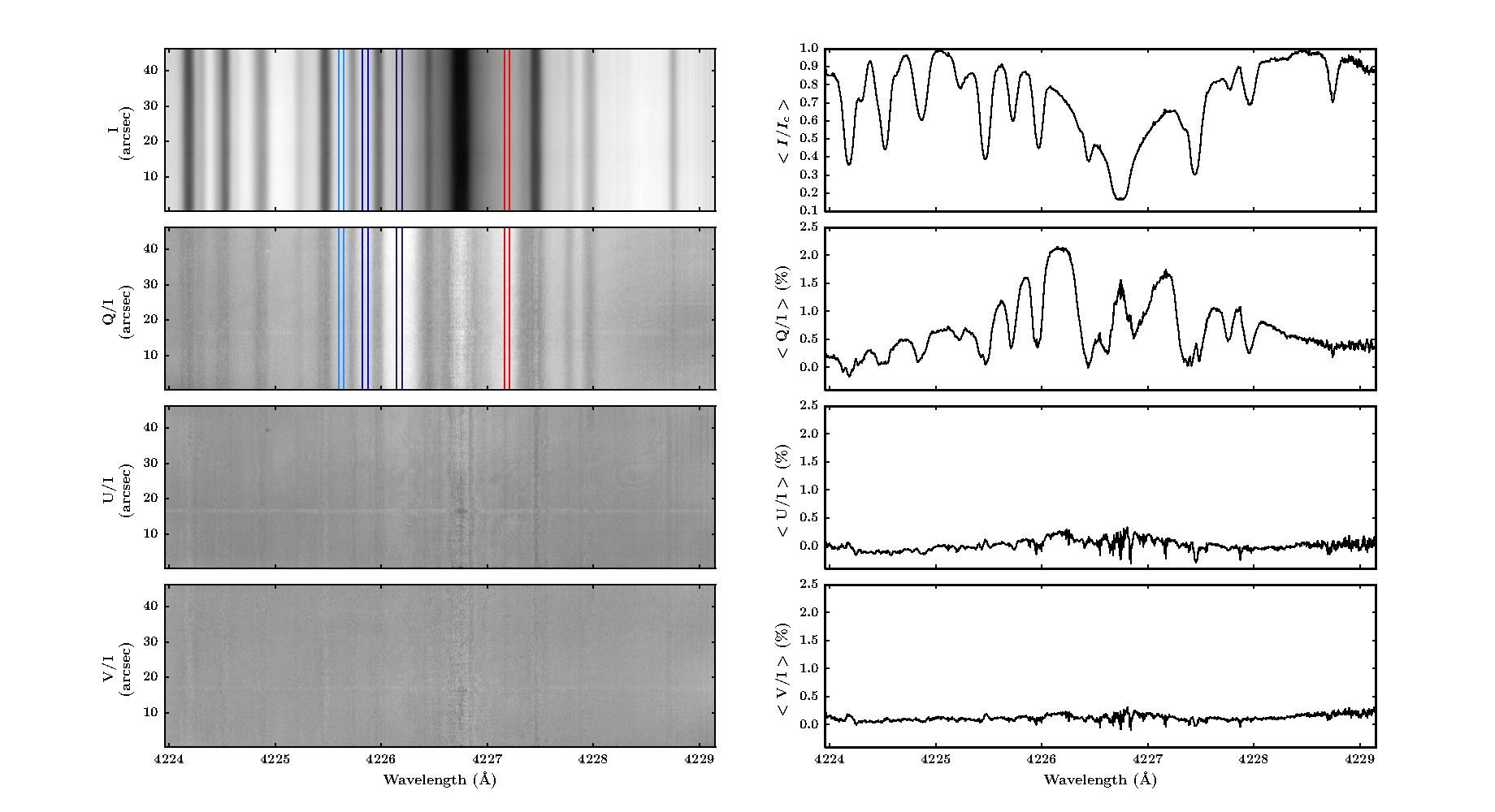}
	\caption{Stokes $I$, $Q/I$, $U/I$, and $V/I$ images 
	(left) and spatially-averaged profiles (right). The spatial average was 
	performed over the entire slit. 
	The intensity is normalized to the continuum.
	The observations were taken at ${\sim}6\arcsec$ from the solar limb, 
	across a quiet region, with the slit oriented parallel to the nearest
	limb. The positive direction for Stokes $Q$ is taken parallel to the limb.
	The wavelength intervals of interest are highlighted in the intensity 
	and $Q/I$ images.} 
\label{fig:m4stoksimageprofiles}
\end{figure*}
\begin{figure}[t!]
\includegraphics[width=0.5\textwidth]{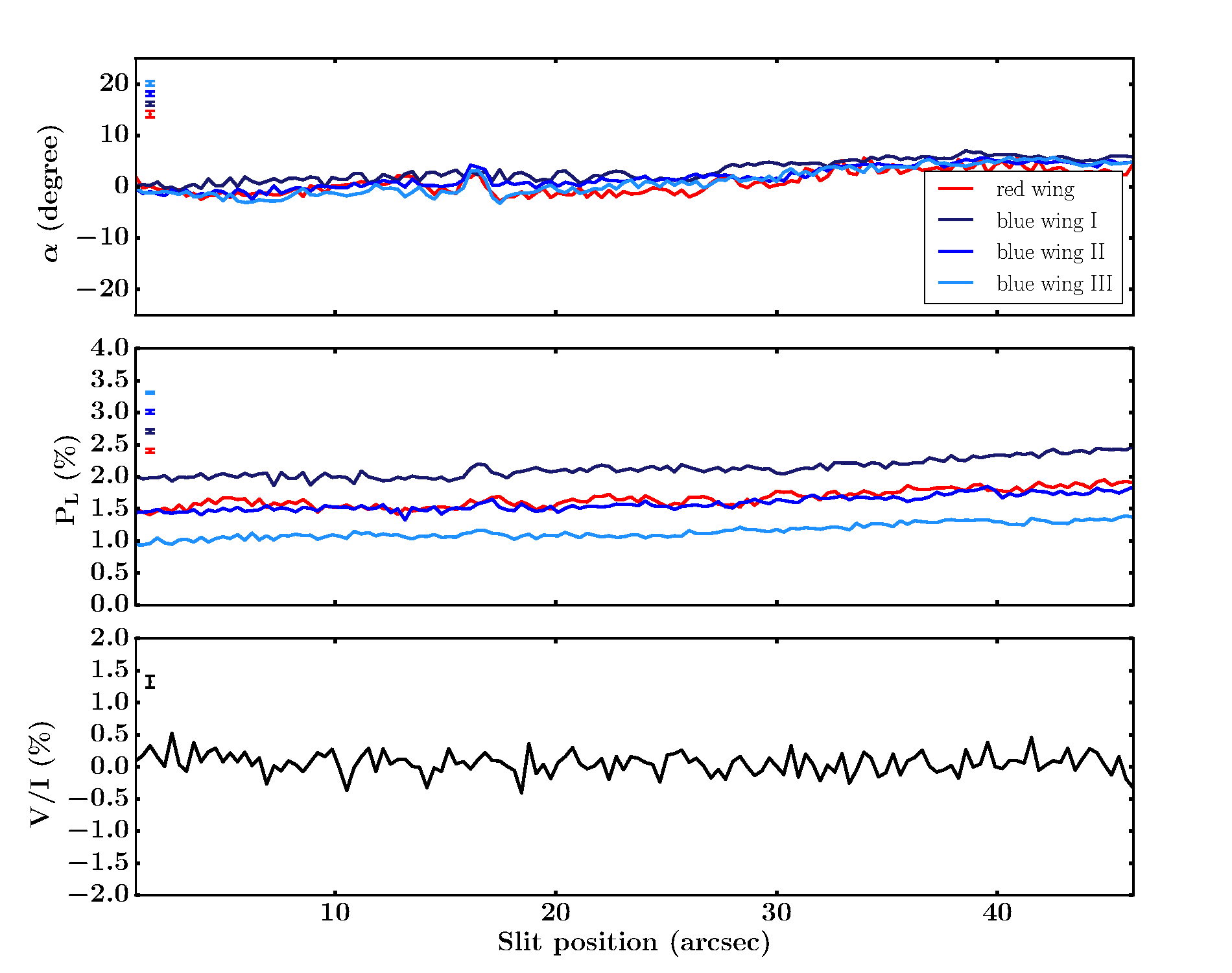}
	\caption{Linear polarization angle (top panel) and total linear 
	polarization fraction (middle panel) along the slit for the 
	considered wavelength intervals (see text and Table~\ref{Table:1}). 
	In the bottom panel, the amplitude of the
	blue peak of the $V/I$ signal in the Fe~{\sc i} 4224.2\,{\AA} line
	is shown as a function of the slit position.
	The values shown in all three panels are obtained from the data presented 
	in Figure~\ref{fig:m4stoksimageprofiles}. 
	The error bars (statistical error) are reported in the upper left part of each panel.}
\label{fig:4227m4gregorpolA}
\end{figure}

\section{Observations at GREGOR} \label{sec:sec4} 
The first observations that we analyze were taken at the GREGOR telescope on June, 24th 2018,
using the ZIMPOL camera \citep[see][]{Ramelli10}, which was installed at the 
GREGOR spectrograph.
The polarimeter analyzer, consisting of a double ferroelectric crystal (FLC) 
modulator followed by a linear polarizer, was mounted in front of the 
spectrograph slit after several folding mirrors; the FLC modulates the 
polarization signals at the frequency of 1\,kHz, allowing to ``freeze'' 
intensity variations due to seeing \citep[see][for technical 
details]{Ramelli14}.

The spectrograph slit covers a solar area of ${\sim}0.3\arcsec$ (width of the slit) times $47\arcsec$ 
(length of the slit).
The ZIMPOL images presented in this work have 140 pixels along the spatial 
direction and 1240 pixels along the spectral direction.
According to the Nyquist criterion, the attainable spatial resolution  
is ${\sim}0.6\arcsec$, while the spectral resolution is of ${\sim}10$\,m\AA .

The GREGOR polarimetric calibration unit \cite[GPU,][]{Hofmann12} is mounted 
at the second focal point (F2) before any folding reflection, in order to 
avoid any significant instrumental polarization produced by the mirrors.
Due to the presence of a pre-filter in the GPU that blocks the 
radiation at all wavelengths below a cutoff just above the Ca~{\sc i} line, 
the polarimetric calibration measurements were performed at around 4240\,{\AA} and not at 
the exact wavelength of the line.
The polarimetric calibration, the dark images and the flat field (obtained by 
moving the telescope randomly around disc center) were recorded within a few minutes 
of the observations \citep [see][for more details]{Dhara19}.

We carried out two observations: one on an active region close to the West 
limb, and one on a quiet region at a diametrically opposed position at the 
East limb. The latter was carried out in order to have an observation in the quiet Sun to be compared 
to the one in the active region.
For scheduling reasons, the observations were carried out around noon, 
which is a comparatively unfavorable time because the althazimuthal system rotates
quite fast. This introduced time-dependent instrumental polarization signals that could
not be completely removed in the data reduction. 
Moreover the seeing was rather poor ($r_{0} < 4$), but still good enough for the purpose of this work.
The slit was placed parallel to the limb, choosing the positive direction for 
Stokes $Q$ to be parallel to the slit direction.
The tip-tilt of the adaptive optics (AO) system \citep{Berkefeld16} was used
to keep the limb distance constant with an uncertainty of $\sim 0.5\arcsec$.
The data of the two observations were corrected using the flat-field and dark 
image, and a Fourier filter was also used to remove the periodic fringes 
originated by both electronics and optics.
Furthermore, because the absolute value of the polarization 
cannot be determined with the instrumental set-up used in this campaign,
the Stokes profiles of the measurements were properly shifted along the 
polarization scale so that in the continuum $Q/I$ matches the theoretical 
value \citep{Gandorfer02} while $U/I$ and $V/I$ are zero.

For each Stokes parameter, the statistical error has been determined as follows. 
A slit portion of 20 pixels, corresponding to a magnetically uniform region, is selected.
Then, at each frequency, the standard deviation is taken over the considered portion. 
The one with the highest occurrence is taken as the statistical error for all frequency points. 
On the other hand, the statistical error associated to the linear polarization fraction and linear polarization angle 
is taken as the average absolute difference between such quantities, 
calculated from the experimental data at the all spatial positions and a smooth function fitting their 
variation along the slit.

\subsection{GREGOR observation on a quiet region: analysis}\label{sec:sec4_1} 
We observed a quiet Sun region close to the East limb, at 
a limb distance of ${\sim}6\arcsec$.
The duration of the measurement was around 8 minutes (240 frames of 
1 second exposure time each, with 1 second of lag between two consecutive frames).
Figure~\ref{fig:m4stoksimageprofiles} shows this measurement and the 
corresponding spatially averaged Stokes profiles. 
The statistical error was determined as described in section \ref{sec:sec4}.
It is 0.09$\%$ for Stokes $Q/I$ and $V/I$ and 0.06$\%$ for Stokes $U/I$.

The Stokes $Q/I$ profile shows the aforementioned triple peak structure, with a sharp peak at the line core and
extended wing lobes; the highest polarization value in the blue lobe reaches ${\sim}2\%$.
We notice the persistence of some spurious effects in the $U/I$ and $V/I$ 
profiles, whose origin can be attributed to the fast changes 
of the instrumental polarization at the time of day at which the 
observation was taken (see section \ref{sec:sec4}).

As discussed in section \ref{sec:sec2}, we focus our attention on four 
wavelength intervals in the wings, which are highlighted in 
Figure~\ref{fig:m4stoksimageprofiles} and whose distances 
from line center are given in Table~\ref{Table:1}. 
\begin{table}[h!]
\centering
\footnotesize
 	\caption{\textbf{Wing wavelength intervals for Figure~\ref{fig:m4stoksimageprofiles}.}}
\begin{tabular}{|c|c|} 
\hline
 Wavelength interval & Distance from line center [{\AA}] \\
\hline  
 bI    &  0.57      \\
 bII   &  0.89       \\
 bIII  &  1.12      \\
 rI    &  0.44       \\
\hline
\end{tabular}
\label{Table:1}
\end{table}

The upper panel of Figure~\ref{fig:4227m4gregorpolA} shows that the linear 
polarization angle is very close to zero along the whole slit for all the
considered wavelengths.
The anomalous peak found between $15\arcsec$ and $17\arcsec$ is of instrumental 
origin, resulting from dust grains in the optical system, close to the focal plane.
This is supported by the bright horizontal stripe found at the same spatial 
position in Figure~\ref{fig:m4stoksimageprofiles}.
Also the total linear polarization fraction does not show 
any significant variation along the slit, at any of the considered wavelengths
(see middle panel of Figure~\ref{fig:4227m4gregorpolA}).
These results are compatible with our hypothesis because we are observing 
a quiet Sun region, where no magnetic fields capable of producing significant MO effects are expected.
The absence of magnetic fields with a significant longitudinal component is 
confirmed by the very weak $V/I$ signals observed in the nearby Fe~{\sc i} 
line (see lower panel of Figure~\ref{fig:4227m4gregorpolA}).

\subsection{GREGOR observation on an active region: analysis}
\label{sec:sec4_2}  
\begin{figure*}[h!]
\centering
\includegraphics[width=1.\textwidth]{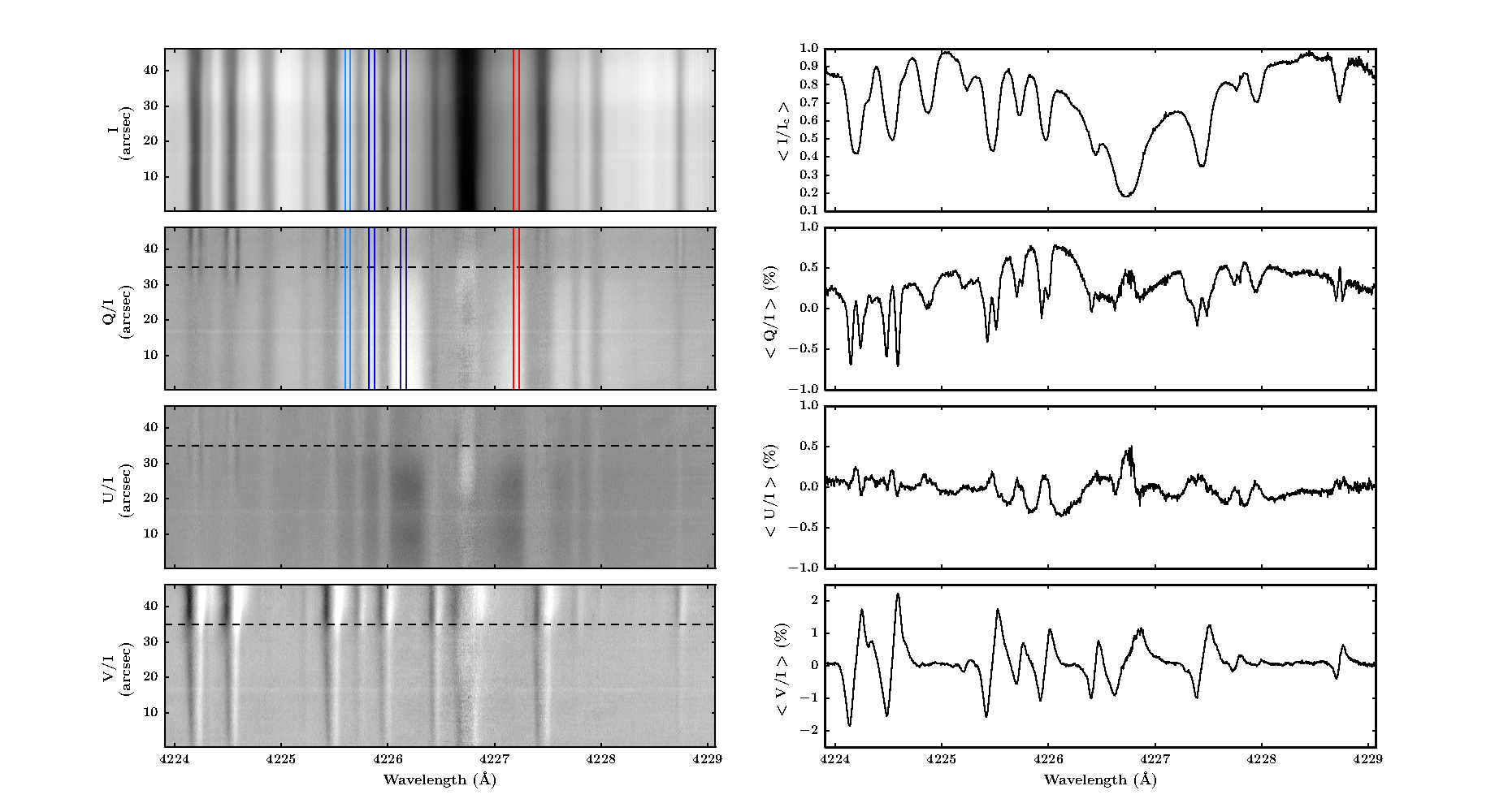}
	\caption{Stokes $I$, $Q/I$, $U/I$, and $V/I$ images 
	(left) and spatially-averaged profiles (right).
	The spatial average was performed over the slit positions from 
	${\sim}35\arcsec$ to $47\arcsec$ (dashed line) where the magnetic field 
	is strongest. The intensity is normalized to the continuum.
	The observations were taken at a limb distance of $4\arcsec$, across a 
	solar region with moderate magnetic activity, with the slit oriented 
	parallel to the nearest limb. The positive direction for Stokes $Q$ is 
	taken parallel to the limb. The wavelength intervals of interest are 
	highlighted in the intensity and $Q/I$ images.} 
\label{fig:m1stoksimageprofiles}
\end{figure*}

At GREGOR, we also observed a region at $4\arcsec$ from the 
West limb with modest magnetic activity, 
with part of the slit placed upon a region with  
moderate activity (see slit portion from ${\sim}35\arcsec$ to 
$47\arcsec$ in Figure~\ref{fig:m1stoksimageprofiles}).
The duration of the measurement was around 9 minutes (284 frames of 1 second 
exposure time each, with 1 second of lag between two consecutive frames).
The statistical error was determined as described in section \ref{sec:sec4}. 
It is 0.08$\%$ for Stokes $Q/I$ and $V/I$ and 0.07$\%$ for Stokes $U/I$.

\begin{figure}[h!]
\includegraphics[width=0.5\textwidth]{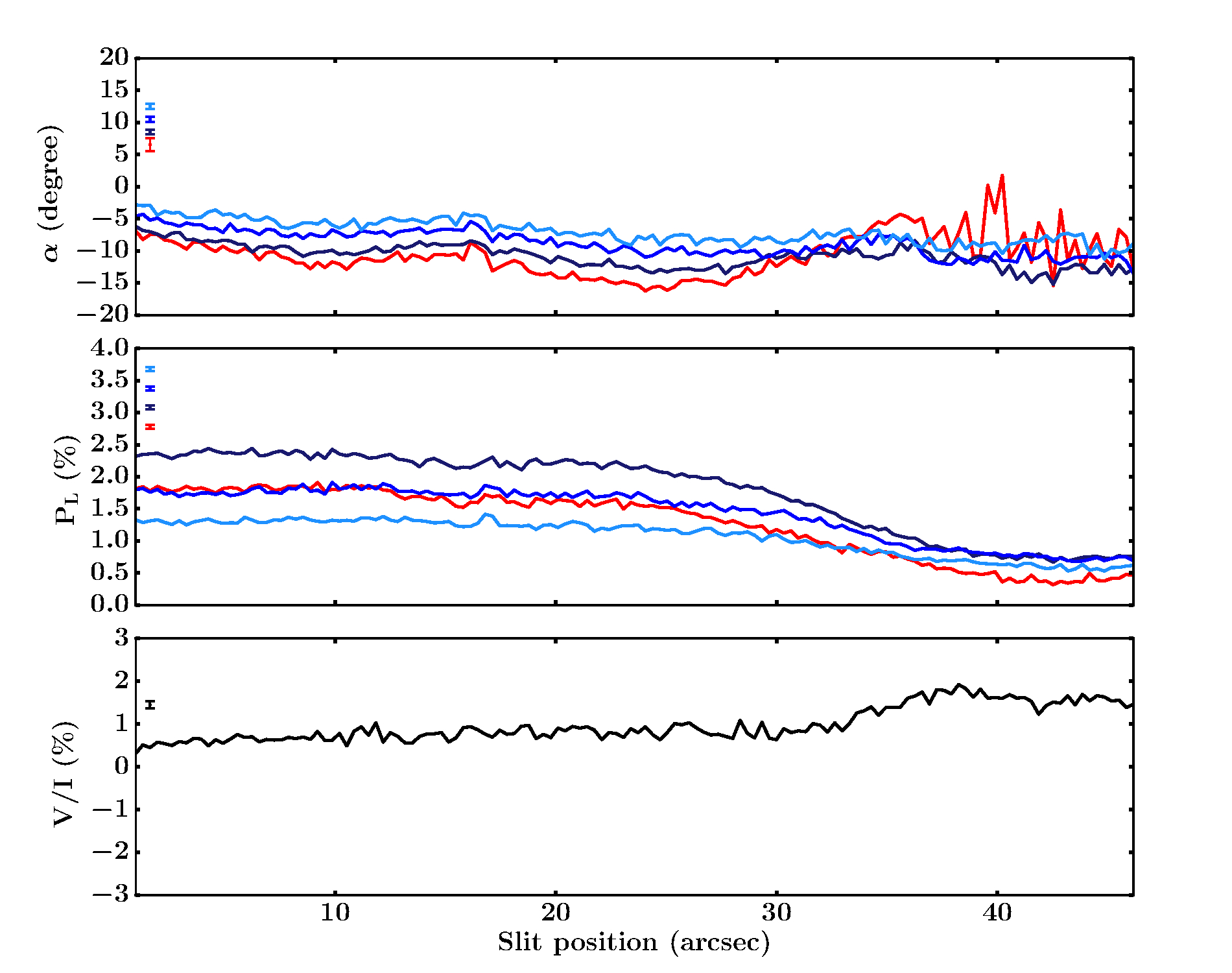}
	\caption{
	Linear polarization angle (top panel) and total linear polarization 
	fraction (middle panel) along the slit for the 
	considered wavelength intervals (see text and Table~\ref{Table:2}). 
	In the bottom panel, the amplitude of the
	blue peak of the $V/I$ signal in the Fe~{\sc i} 4224.2\,{\AA} line
	is shown as a function of the slit position.
	The values shown in all three panels are obtained from the data presented 
	in Figure~\ref{fig:m1stoksimageprofiles}. 
	The error bars (statistical error) are reported in the upper left part of 
	each panel.}
\label{fig:4227m1gregorpolA}
\end{figure}
Figure~\ref{fig:m1stoksimageprofiles} shows the Stokes images (left panels) 
and the corresponding spatially averaged profiles (right panels), 
for which the average was performed within
the region where the magnetic field is strongest (i.e., over the 
slit portion from ${\sim}35\arcsec$ to $47\arcsec$).
In this region, Stokes $Q/I$ shows again a triplet peak structure.
However the line core peak, in comparison to the previous measurement (see 
Figure~\ref{fig:m4stoksimageprofiles}), is clearly depolarized through the Hanle effect, which also induces 
significant $U/I$ line-core signals. 
In the line wings prominent $U/I$ signals are found along the entire slit. 
A significant depolarization of both $Q/I$ and $U/I$ wing signals is found in the region between ${\sim}35\arcsec$ and $47\arcsec$.
Indeed the blue wing maximum of the spatially averaged $Q/I$ signal within this region is significantly smaller than in 
the previous measurement, being in this case of $\sim 0.7\%$.
In the Stokes $V/I$ and $Q/I$ images, the typical patterns of the longitudinal and transverse
Zeeman effect, respectively, can be observed.

\begin{table}[h!]
\centering
\footnotesize
 	\caption{\textbf{Wing wavelength intervals for Figure~\ref{fig:m1stoksimageprofiles}.}}
\begin{tabular}{|c|c|} 
\hline
 Wavelength interval & Distance from line center [{\AA}] \\
\hline  
 bI    &  0.59   \\
 bII   &  0.89   \\
 bIII  &  1.11   \\
 rI    &  0.47    \\
\hline
\end{tabular}
\label{Table:2}
\end{table}
The line-center distance of the wavelength intervals of interest for this 
observation, as defined in section \ref{sec:sec2}, are shown in 
Table~\ref{Table:2}. 
As can be seen in Figure~\ref{fig:4227m1gregorpolA}, the polarization angle differs appreciably from zero
along the whole slit at all the considered wing wavelengths. 
This rotation of the plane of linear polarization is compatible with the impact of MO effects produced 
by a longitudinal magnetic field, whose presence is revealed by the Stokes $V/I$ signal 
in the photospheric Fe~{\sc i} 4224.2\,\AA\ line. 
An increase of the amplitude of this $V/I$ signal can be clearly observed between ${\sim}35\arcsec$ and $47\arcsec$,
confirming the presence of stronger longitudinal fields in this region (see the bottom panel of Figure~\ref{fig:4227m1gregorpolA}).
It is interesting to observe that while $P_L$ shows a clear decrease
in this region, the variation of the polarization angle is less pronounced.
This behavior can be interpreted observing that the anisotropy of the radiation
field and other properties of the atmospheric plasma (e.g., the density of neutral perturbers) have a direct impact on $P_L$,
whereas they influence $\alpha$ only through its sensitivity to the magnetic field (Capozzi et al. in preparation).
The observed variation of $P_L$ could thus be mainly ascribed to a variation of the properties of the atmospheric 
plasma in the region between ${\sim}35\arcsec$ and $47\arcsec$, rather than to the presence of stronger fields via MO effects.

\section{Observation at IRSOL} \label{sec:sec5}
Another set of spectropolarimetric observations of the Ca~{\sc i} 
line at 4227\,\AA\ was performed with the ZIMPOL camera at 
IRSOL on April 19, 2019.
The instrumental set-up was similar to the one used during the GREGOR campaign 
(section \ref{sec:sec4}), but the polarization modulation was done 
with a photoelastic modulator (PEM), which modulates the signals with a 
frequency of 42\,kHz, allowing us to minimize spurious effects induced by 
intensity variations due to the seeing.

The spectrograph slit covered a solar area of $0.5\arcsec$ (width of the slit) 
times $180\arcsec$ (length of the slit).
Recalling that the ZIMPOL image has 140 pixels in the spatial direction and 
1240 pixels in the spectral direction, the spatial sampling is 
${\sim}1.47\arcsec$/pixel, and the spectral resolution is 
${\sim}10$\,m{\AA}.
The polarimetric calibration, the dark image and the flat field image (obtained 
by moving the telescope randomly around disc center) were all recorded within 
few minutes of the observations.
The data was corrected for flat-field and dark images, and a Fourier filter 
was used.
Since the polarimetric calibration unit at IRSOL is mounted after two folding 
mirrors, some cross-talk is unavoidable. 
The cross-talk correction was made by performing a rotation around the $U$ and 
$Q$ axes of the Poincar\'{e} sphere until the $V/I$ cross-talk signal was 
minimized.
In the observation that we have analyzed, the cross-talk from 
$V/I$ to $Q/I$ was corrected by a 6$^{\circ}$ rotation around the $U$ axis, and
the cross-talk from $V/I$ to $U/I$ was corrected by a rotation around the $Q$ 
axis with an angle of 5.5$^{\circ}$.

By using a rotating glass tilt plate based on an elaboration of the slit jaw image, the distance of the limb to the spectrograph slit
could be kept constant within an accuracy of $\sim 1\arcsec$.
Given the length of the slit of the IRSOL telescope we also corrected for the effects of the curvature of the limb \citep{Bianda11}.

\subsection{Analysis}\label{sec:sec5_1}
\begin{figure*}[h!]
\centering
\includegraphics[width=1.\textwidth]{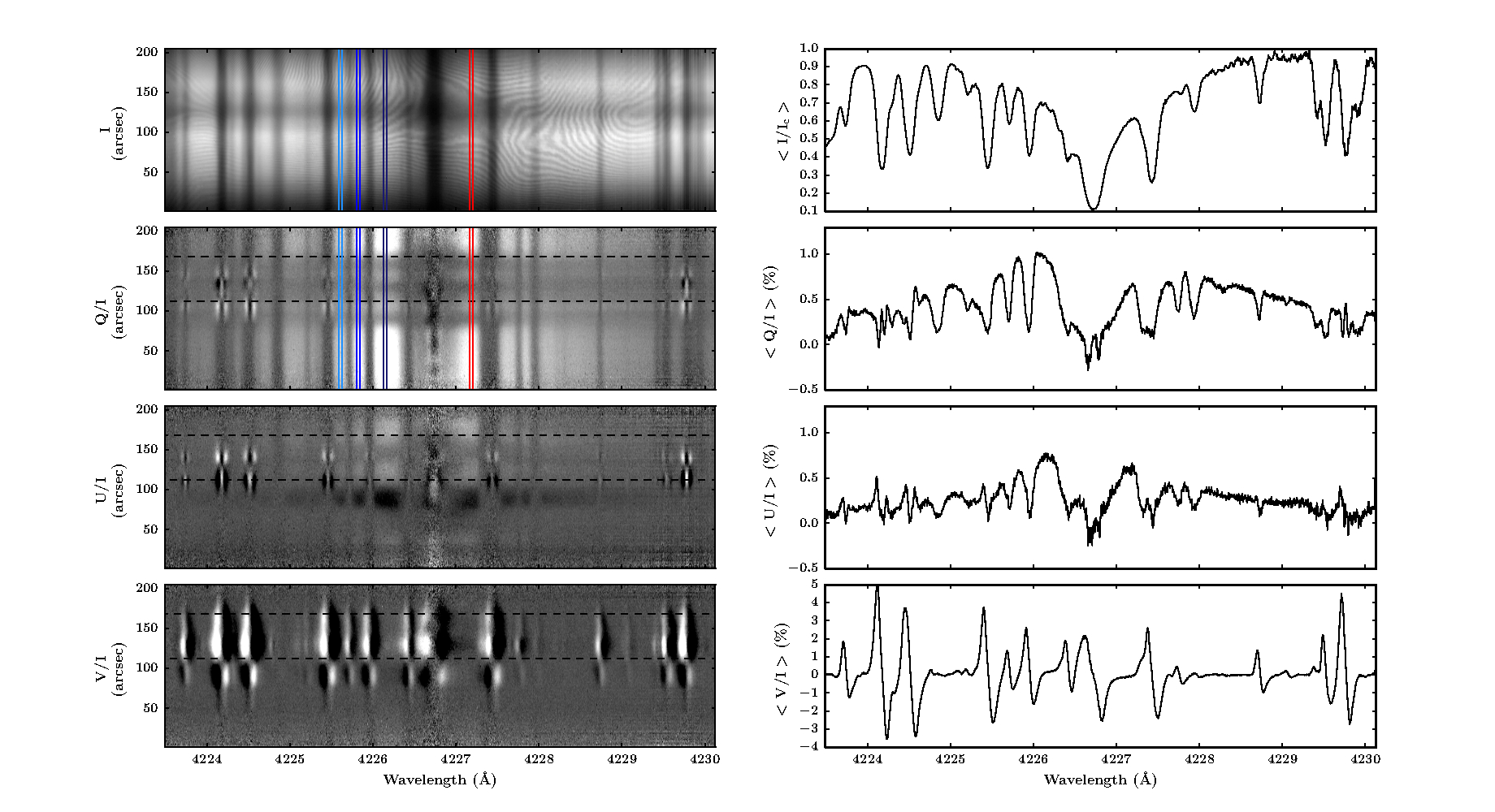}
	\caption{Stokes $I$, $Q/I$, $U/I$, and $V/I$ images 
	(left) and spatially-averaged profiles (right). 
	The spatial average was performed over the slit positions from 
	${\sim}110\arcsec$ to $170\arcsec$ (dashed line) where the magnetic field
	is strongest. The intensity is normalized to the continuum. 
	The observation was taken at a limb distance of $5\arcsec$, across a solar 
	region with stronger magnetic activity, with the slit oriented 
	parallel to the nearest limb. The positive direction for Stokes $Q$ is taken parallel to the limb. 
	The wavelength intervals of interest are highlighted in the intensity 
	and $Q/I$ images.} 
\label{fig:m4irsolstoksimageprofiles}
\end{figure*}

We observed a region at $5\arcsec$ from the West limb with moderate magnetic 
activity.
The duration of the measurement was around 7 minutes (200 frames of 
1 second exposure time each, with 1 second of lag between two consecutive frames).
The slit was placed parallel to the nearest limb, and the reference direction 
for positive Stokes $Q$ was chosen along the slit.
The left panels of Figure~\ref{fig:m4irsolstoksimageprofiles} show the Stokes images while the right panels show the 
corresponding profiles spatially averaged over the slit portion between ${\sim}110\arcsec$ and $170\arcsec$
(i.e. over a region in which we have a strong magnetic field with a fixed polarity).
The central peak of this $Q/I$ profile is strongly depolarized through the Hanle 
effect. The maximum value of $Q/I$ in the blue wing reaches ${\sim}1\%$ and prominent $U/I$ signals are also found.
In the region from ${\sim50}\arcsec$ to ${\sim}170\arcsec$, 
one can recognize the typical patterns of the longitudinal Zeeman effect 
in the Stokes $V/I$ image, as well as transverse Zeeman signals in Stokes $Q/I$ and $U/I$.
The statistical error was determined as described in section \ref{sec:sec4}. 
It is 0.09$\%$ for Stokes $Q/I$ and 0.11$\%$ for Stokes $U/I$ and $V/I$.

\begin{figure}[h!]
\includegraphics[width=0.5\textwidth]{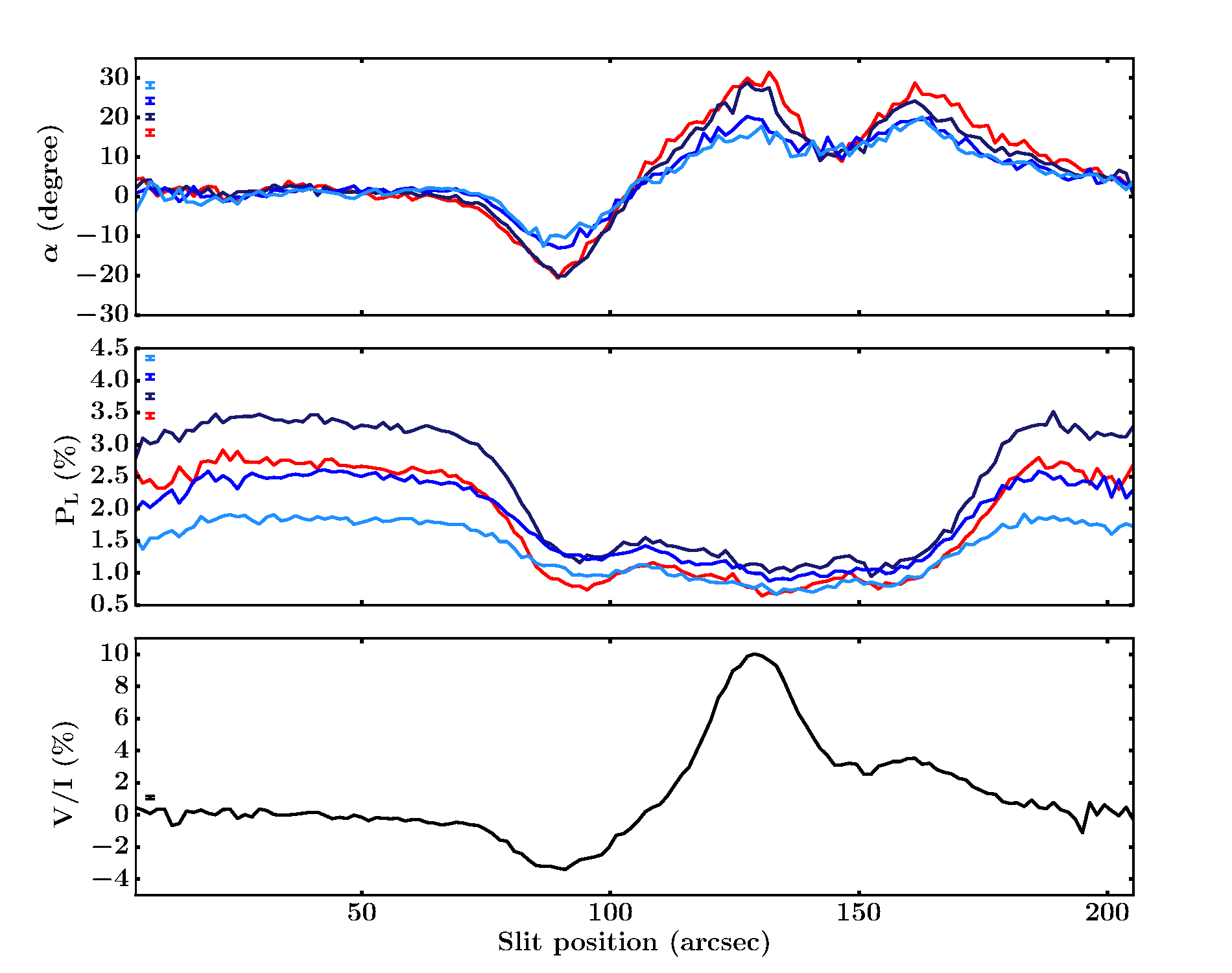}
	\caption{Linear polarization angle (top panel) and total linear polarization 
	fraction (middle panel) along the slit for the 
	considered wavelength intervals (see text and Table~\ref{Table:3}). 
	In the bottom panel, the amplitude of the
	blue peak of the $V/I$ signal in the Fe~{\sc i} 4224.2\,{\AA} line
        is shown as a function of the slit position.
	These values are obtained from the data presented in 
	Figure \ref{fig:m4irsolstoksimageprofiles}. 
	The error bars (statistical error) are reported in the upper left part of each panel.}
\label{fig:4227m4irsolpolA}
\end{figure}

In Figure~\ref{fig:4227m4irsolpolA} the behavior of $\alpha$, $P_L$ and $V/I$ along the slit is shown.
The line-center distance of the wavelength intervals of interest considered in this observation, as 
defined in section \ref{sec:sec2}, are shown in Table~\ref{Table:3}.
\begin{table}[h!]
\centering
\footnotesize
 	\caption{\textbf{Wing wavelength intervals for Figure~\ref{fig:m4irsolstoksimageprofiles}.}}
\begin{tabular}{|c|c|} 
\hline
 Wavelength interval & Distance from line center [{\AA}] \\
\hline  
 bI    &  0.58  \\
 bII   &  0.90  \\
 bIII  &  1.11  \\
 rI    &  0.46  \\
\hline
\end{tabular}
\label{Table:3}
\end{table}

Similarly to what is reported in Sect.~\ref{sec:sec4_2}, a clear change in the behavior of both $P_L$ and 
$\alpha$ can be noticed when going into the more magnetized region, between ${\sim}70\arcsec$ and ${\sim}170\arcsec$.
The behavior of $\alpha$ closely follows the variations of the $V/I$ signal, including the change of sign in the region around 
$100\arcsec$, strongly suggesting the impact of MO effects.
On the other hand, $P_L$ is found to decrease sharply in this region, but it presents no clear variations within it. 
This could again be an indication that the total linear polarization fraction is mainly determined by the anisotropy of the 
radiation field and other non-magnetic properties of the plasma (such as the rate of elastic collisions), which we may expect not to vary 
significantly within the magnetized region.
Still focusing on the region around $100\arcsec$, we may also propose an alternative
explanation (compatible with the previous one). Within this region, one could expect the
presence of magnetic fields of opposite polarities below the spatial resolution of the observation,
leading to cancellations in both $\alpha$ and $V/I$ (in agreement with their vanishing values found around $100\arcsec$).
It is an interesting consequence of MO effects that, if magnetic fields are present such that their orientation changes at 
scales below the resolution element but above the mean free path of the line's photons, they would still contribute to
depolarizing $P_L$ while at the same time cancellations would occur in $\alpha$ \citep[see Appendix A in] []{Alsina18}. 
Discriminating between these two scenarios is outside the scope of this work, but these observational results already provide 
tantalizing hints about the diagnostic potential of the MO effects.

\section{Discussion and conclusions}\label{sec:sec6}
In the present paper we have analyzed different spectropolarimetric 
observations of the Ca~{\sc i} 4227\,{\AA} line, obtained 
using the ZIMPOL camera with both the GREGOR and IRSOL telescopes.
Spatial variations of the $Q/I$ scattering polarization wing signals
as well as the appearance of $U/I$ wing signals, presenting similar spatial fluctuations,
have been clearly detected.
Comparison with the $V/I$ signals produced by the Zeeman effect in the
nearby Fe~{\sc i} 4224.2\,\AA~ line clearly indicate that such 
spatial variations are of magnetic origin, thus supporting 
the conclusion that they represent measurable signatures of MO effects operating on this line, as 
theoretically predicted in \cite{Alsina18}.

To make further advances in this line of investigation it will be 
of interest to have more quantitative information of the magnetic field present in the observed region. 
In this regard the use of advanced spatially coherent inversion techniques
\citep[e.g.][]{AsensioRamosdelaCruzRodriguez15} will be highly 
beneficial in acquiring much needed information.

The results presented in this work also provide some hints on
the potential of MO effects as a novel tool for magnetic field diagnostics, complementary to the Zeeman and Hanle effects.
This motivates investigating the impact of MO effects on this line in greater depth.
In a forthcoming publication, we will take a numerical approach, based on 
the solution of the non-LTE radiative transfer problem, in order to 
characterize the sensitivity of the four Stokes profiles of this line to 
magnetic fields, accounting for the joint action of the Hanle, Zeeman, and MO 
effects.

\begin{acknowledgements}
	This research work was financed by the Swiss National 
	Science Foundation (SNF) through grants 200020\_169418
	and 200020\_184952.
	E.A.B. and L.B. gratefully acknowledge financial support 
	by SNF through grants 200021\_175997 and CRSII5\_180238.
	IRSOL is supported by the Swiss Confederation (SEFRI), Canton Ticino, 
	the city of Locarno and the local municipalities. 
	The 1.5-meter GREGOR solar telescope was built by a German consortium under 
	the leadership of the Kiepenheuer-Institut fur Sonnenphysik in Freiburg 
	with the Leibniz-Institut f\"ur Astrophysik Potsdam, the Institut 
	f\"ur Astrophysik G\"ottingen, and the Max-Planck-Institut f\"ur 
	Sonnensystem forschung in G\"ottingen as partners, and with contributions 
	by the Instituto de Astrof\'{i}sica de Canarias and the Astronomical Institute 
	of the Academy of Sciences of the Czech Republic.
\end{acknowledgements}

\bibliographystyle{aa}
\bibliography{calcium_aa}

\end{document}